\documentclass[prb,twocolumn,superscriptaddress,showpacs,floatfix]{revtex4}
\usepackage{graphicx}
\usepackage{amssymb}
\usepackage{amsmath}
\usepackage{citesort}
\usepackage{color}
\usepackage{times,mathptm}

\begin{document}

\title{Manifestations of fine features of the density of states\\ in
the transport properties of $\rm \mathbf{KOs_2O_6}$}

\author{A. Akrap}
\affiliation{Institut de Physique de la mati\`{e}re complexe,
EPFL, CH-1015 Lausanne, Switzerland}

\author{E. Tuti\v s}
\affiliation{Institut za fiziku, Bijeni\v{c}ka cesta 46, Zagreb,
Croatia}

\author{S. M. Kazakov}
\affiliation{Laboratory for Solid State Physics, ETH Z\"urich,
8093 Z\"urich, Switzerland} \affiliation{Department of Chemistry,
Moscow State University, 119899 Moscow, Russia }

\author{N. D. Zhigadlo}
\author{J. Karpinski}
\affiliation{Laboratory for Solid State Physics, ETH Z\"urich,
8093 Z\"urich, Switzerland}

\author{L. Forr\'o}
\affiliation{Institut de Physique de la mati\`{e}re complexe,
EPFL, CH-1015 Lausanne, Switzerland}

\begin{abstract}
We performed high pressure transport measurements on high quality
single crystals of $\rm KOs_2O_6$, a $\beta$-pyrochlore
superconductor. While the resistivity at high temperatures might
approach saturation, there is no sign of saturation at low
temperatures, down to the superconducting phase. The anomalous
resistivity is accompanied by a non-metallic behavior in the
thermoelectric power (TEP) up to temperatures of at least
$700\rm\,K$, which also exhibits a broad hump with a maximum at
$60\rm\,K$. The pressure influences mostly the low energy
electronic excitations. A simple band model based on enhanced
density of states (DOS) in a narrow window around the Fermi energy
($E\rm_F$) explains the main features of this unconventional
behavior in the transport coefficients, and its evolution under
pressure.
\end{abstract}


\maketitle

Recent discovery\cite{KOSOdiscSC,KOSOunprecedentedSC} of
superconductivity at $T\rm_c$ of $9.6\rm\,K$ in
pyrochlore-structured $\rm KOs_2O_6$ has strengthened the
viewpoint that magnetic frustration may have considerable role in
promoting superconductivity.\cite{AokiREV} Pyrochlore structure is
known to impose frustration on magnetic ordering in several
systems with localized spins, although the role of frustration in
electronically itinerant pyrochlore systems is presently unclear.
$\rm KOs_2O_6$ has the highest superconducting transition
temperature among the pyrochlore-structured compounds, and is
therefore of particular interest. However, there are indications
that geometrical frustration may not be the most important factor
for the physics of this compound. $\rm OsO_6$ octahedrons,
positioned at the nodes of a pyrochlore lattice, form large cages
where potassium atoms move in an anharmonic
potential.\cite{PickettRattling} This "rattling motion" of the K
ions appears to have significant influence on the physical
properties in this material. The electronically identical Rb and
Cs compounds have much lower $T\rm_c$'s,\cite{RbSC,CsSC} and
simultaneously, this rattling motion is significantly less
pronounced. Another remarkable property of $\rm KOsO_2$ is the
large value of the Sommerfeld constant found in the specific heat
measurements.\cite{JapSpecHeat,Markus} The Sommerfeld coefficient
implies a density of states (DOS) at the Fermi energy ($E_F$) an
order of magnitude higher then the one predicted by the band
structure calculations.\cite{PickettRattling,SanizBAND} In this
paper we argue that the high DOS at the Fermi level dominates the
shape of the electronic transport properties in the normal state
over a wide range of temperatures and pressures.

Interestingly, frustration, the rattling of the K ions and the
high DOS at $E\rm_F$ in a broad sense relate $\rm KOs_2O_6$ to the
high temperature superconducting cuprates. There, the magnetic
frustration is related to missing spins introduced by doping,
often at the expense of the non-stoichiometric crystal structure.
The strongly anharmonic phonon modes are also often encountered in
the cuprates and are sometimes regarded as important for
superconductivity.\cite{egApexOxigen} Lastly, the band structure
calculations and photoemission measurements suggest a van Hove
enhancement in the DOS in the vicinity of $E\rm_F$.\cite{vanHove}
Similarly to the copper oxide superconductors, the normal state
properties of $\rm KOs_2O_6$ may prove more intriguing than those
of the superconducting phase.

Application of pressure on $\rm KOs_2O_6$ may tune the coupling
constants of the electron-electron and electron-phonon
interaction, change the size of the rattling cage, or widen the
bands. We have measured the high pressure behavior of the
transport coefficients, resistivity $\rho(T)$ and the
thermoelectric power (TEP) $S(T)$ on high quality single crystals
of $\rm KOs_2O_6$. Previously reported high pressure resistivity
measurements were done on polycrystalline samples,\cite{hpPRL}
which made it difficult to separate the intrinsic behavior from
the grain-size effects. Focusing on the anomalous normal state
rather than on the superconducting phase, we also investigated the
transport properties on a much finer pressure scale than
previously reported.\cite{hpPRL} Additionally, we have measured
the ambient pressure transport coefficients up to high
temperatures, reaching $700\rm\,K$. The TEP data indicates a
behavior unusual for ordinary metals stretching up to at least
$700 \rm\,K$, which links to the unconventional resistivity
$\rho(T)$. We argue that this may be understood as a result of
high electronic DOS, confined to a rather narrow energy window
around $E\rm_F$. Such a DOS enhancement, which is indicated by the
large Sommerfeld coefficient, is probably related to the rattling
mode, as our high pressure results suggest.

The growth of single crystals of $\rm KOs_2O_6$ is described in
the work of G. Schuck \textit{et al}.\cite{crystalgrowth}
Resistivity was measured by a standard four-point method on a
crystal of approximate dimensions $0.2\times 0.2\times 0.2\,\rm
mm^3$. For the TEP measurement, which is insensitive to the grain
size, we used a single-crystal conglomerate approximately $1\rm\,
mm$ long. The sample was attached onto a ceramic surface, which
was heated by a small metallic heater. The temperature gradient
was determined by a differential thermocouple and the contribution
of the golden leads to the TEP was subtracted. The measurements
were carried out in a clamped piston pressure cell, with an $\rm
InSb$ pressure gauge determining the pressure \textit{in situ}.
The applied pressure was hydrostatic, and the maximum pressure
reached was $2.3\rm\,GPa$.

\begin{figure}[h]
\centering
\includegraphics[totalheight=6.5cm]{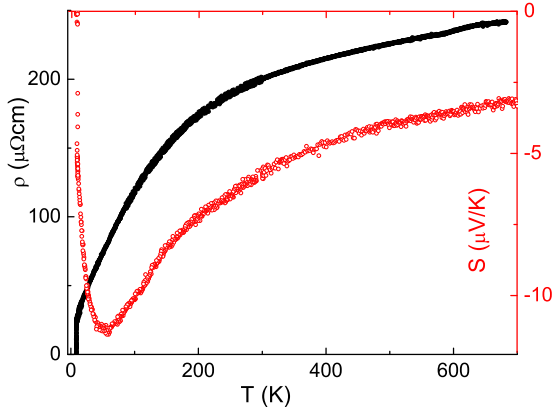}
\caption{(Color online) The wide-range temperature dependencies of
resistivity (left) and thermoelectric power (right) at ambient
pressure. \label{rhoTEP}}\end{figure}

The ambient pressure resistivity and TEP measured in the
temperature range from $4.2$ to $700\rm\,K$ are shown in Figure
\ref{rhoTEP}. The quality of the crystals is reflected in a
comparatively high residual resistivity ratio (RRR) of 15. Since
there is no saturation in resistivity at low temperatures, under
residual resistivity, $\rho_0$, we refer here to the value of the
resistivity right above the superconducting transition. At
temperatures above $200\rm\,K$, $\rho(T)$ grows, exhibiting
neither a strong increase which we would have in case of
scattering on phonons, nor a saturation, which was seen, for
instance, in the pyrochlore superconductor $\rm
Cd_2Re_2O_7$.\cite{NevenCdReO} A plausible reason for the absence
of saturation immediately above $200\rm\,K$ is the still large
mean free path. It was estimated to be of the order of ten lattice
constants from the characteristic value of the Fermi velocity
$v_F$ and the DOS obtained in band structure
calculation.\cite{Markus,SanizBAND} At low temperatures, there is
a strong downward curvature in resistivity below $200\rm\,K$.
However, contrary to what was reported previously\cite{hpPRL}, no
concave behavior, such as $\rho_0+AT^2$, is observed in
resistivity at low temperatures, even down to the superconducting
transition temperature $T\rm_c$.

\begin{figure}[ht] \centering
\includegraphics[totalheight=6.5cm]{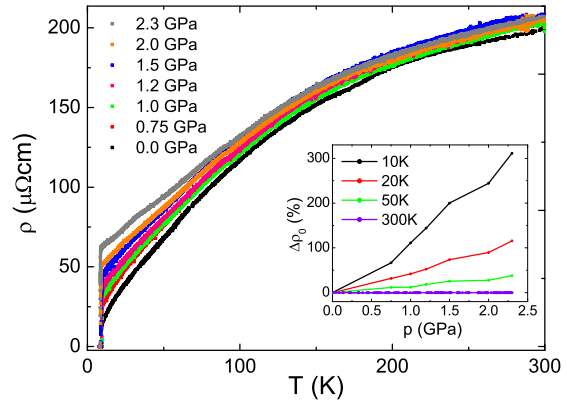}
\caption{(Color online) The temperature dependence of resistivity
for various pressures. The lowest curve is the ambient pressure
resistivity (black). As pressure is applied, the resistivity
increases. The inset shows the pressure dependence of the relative
change of the resistivity with respect to its value at ambient
pressure, for various temperatures. \label{rho}}\end{figure}

The pressure evolution of the resistivity is shown in Figure
\ref{rho}. These are the first reported high pressure measurements
on single crystals of $\rm KOs_2O_6$. At high temperatures the
resistivity is not considerably influenced by the pressure.
However, important changes start to happen in the low temperature
part. The low temperature resistivity at the highest pressure of
$2.3\rm\,GPa$ increases by $300\%$ (inset of Figure \ref{rho}),
which is reflected in the drop of RRR from its ambient pressure
value of 15 to a modest 3.5. Such an increase in the residual
resistivity $\rho_0$ is anomalous. It asserts that pressure
affects mostly the low energy electronic excitations.

\begin{figure}[ht]
\centering
\includegraphics[totalheight=6.5cm]{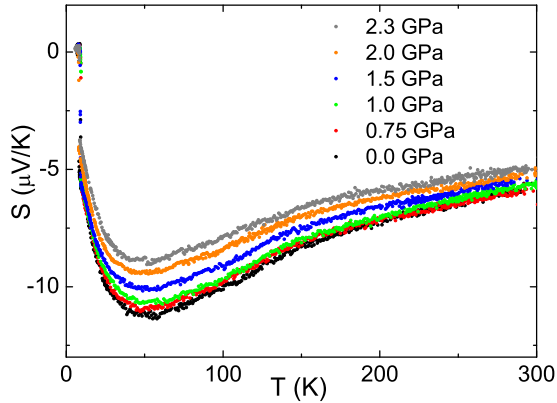}
\caption{(Color online) The temperature dependence of the
thermoelectric power is shown at different pressures. This
transport coefficient decreases in the absolute value as pressure
is applied. \label{TEP}}\end{figure}

The temperature dependence of the TEP up to $700\rm\,K$ is shown
in Figure \ref{rhoTEP}. In the whole temperature range the TEP is
negative. The most prominent feature is a strong peak around
$60\rm\,K$. As temperature is increased further, the TEP drops
precipitously. With the application of pressure, the absolute
value of TEP is reduced, as shown in Figure \ref{TEP}. The largest
changes happen around the maximum, although the position of the
maximum does not shift. Again, the high temperature part of TEP is
much less affected.

There are several reasons to eliminate the interpretation of the
maximum in the TEP as a consequence of a conventional phonon drag.
Even at temperatures as high as $700\rm\,K$ the TEP does not
recover normal metallic behavior, marked by a linear temperature
dependence.\cite{Goodenough} The temperature dependence of
resistivity below $60\rm\,K$ shows no usual signs of the
scattering of electrons on acoustic phonons, as it is convex in
the whole temperature range. In addition, that part of resistivity
strongly depends on pressure, but in the opposite sense to what is
expected if the velocity of the acoustic phonons increases with
pressure. Finally, the maximum value of the TEP decreases with
pressure, contrary to what one would expect if the coupling to
acoustic phonons increased under pressure, as the rise in
resistivity may suggest. The conventional phonon drag being
eliminated, the observed TEP may be only described as an anomalous
electronic contribution. The resistivity and the TEP measurements
together signal the unconventional transport in $\rm KOs_2O_6$.

The presented data accentuate the importance of the low energy
electron dynamics in $\rm KOs_2O_6$. In what follows we will
illustrate that the basic physics of the transport in the normal
state of $\rm KOs_2O_6$ may be understood within a simple
fermionic model with a marked DOS enhancement in the narrow window
of energies. This enhancement is indicated by the Sommerfeld
coefficient $\gamma$ of $75-110\rm\,mJ/(K^2 mol)$ determined from
the specific heat measurements. The Sommerfeld coefficient appears
to be an order of magnitude higher than suggested by band
structure calculations. The increase of the specific heat in
applied magnetic field in the superconducting state\cite{Markus}
demonstrates that the effect is electronic. The additional
contribution to the electronic DOS is likely to be important in a
narrow energy window of about $1\rm\,eV$ around the $E\rm_F$, as
the band structure may be expected to be basically correct for the
$\rm eV$-energy scale. The measurement of TEP suggests that the
half-width of this window is of the order of hundred Kelvin. The
rattling of the K ions is the most suggestive source of the DOS
enhancement. The experimental evidence for this is the increase in
$\gamma$ as we move along the alkali atoms in the isoelectronic $A
\rm Os_2O_6$ series\cite{CsRbSpecHeat} from large Cs, whose
excursions from the equilibrium position are the smallest, over
Rb, to the tiny K, which can rattle the most. On the theoretical
side, it has been pointed out in several recent papers that the
electron scattering on a single rattling mode may be similar to
the Kondo scattering on magnetic impurities, separately for each
spin channel.\cite{KondoRattler} Extended to a crystal where
rattlers reoccur periodically in space, an enhancement of the DOS
around $E\rm_F$ is to be expected, drawing parallels to the heavy
fermion systems.

\begin{figure}[h]
\centering
\includegraphics[totalheight=10cm]{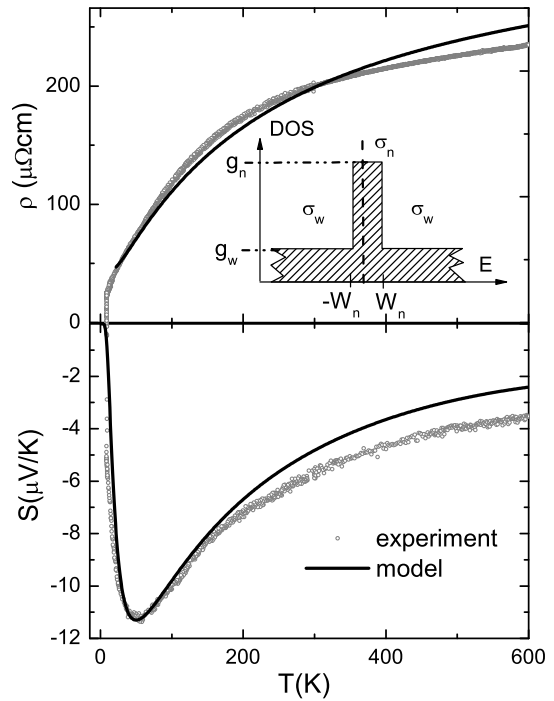}
\caption{The ambient pressure resistivity (top) and TEP (bottom)
fitted by the model depicted in the top panel inset. In the DOS
model, the dashed vertical line sets the position of the band
filling at $T=0$, the values of the DOS in different regions are
indicated by $g\rm_w$ and $g\rm_n$, and $\sigma\rm_w$ and
$\sigma\rm_n$ are the respective conductivity parameters. The
origin of the energy scale is set to the center of the enhanced
part of the DOS. \label{model}}\end{figure}

The simplest fermionic model of the electronic spectrum based on
the observations above is schematically introduced in the inset of
Figure \ref{model}. It consists of a narrow window of the enhanced
DOS around $E\rm_F$ and a wider spectrum where DOS is of the order
of the one given by the band structure calculations. To calculate
the transport coefficients, one needs to evaluate $\sigma(E)=
(e^2/3)\,v(E)^2\,\tau(E)$, the quantity  that includes the
characteristic velocity and relaxation time at a given energy. We
parameterize $\sigma(E)$ for the narrow and wide parts of the
energy spectrum by $\sigma{\rm_{n}}$ and $\sigma{\rm_{w}}$, where
$\sigma{\rm_{n,w}} = (e^2/3) \,v{\rm_{n,w}}^2 \,\tau{\rm_{n,w}}$.
The resistivity and TEP are given by the usual band transport
formulae:\cite{Ziman}
\begin{equation}\label{sigmaTEP}\begin{split}
\sigma&=\int dE\,g(E)\,\sigma(E) \left(-\frac{\partial
f_0 (E,T)}{\partial E} \right) \\
S&=-\frac{e}{T\sigma}\int dE\,g(E)\,\sigma(E) (E-\mu)
\end{split} \end{equation}
where $g(E)$ stands for the DOS and has values $g_n$ and $g_w$,
for the narrow enhanced part and the wings respectively, and
$f_0(E,T)$ refers to the Fermi function. The model is simplistic
to the extent that no implicit temperature dependencies of the
parameters are assumed. In such an approach, the only source of
the temperature dependence of the transport quantities comes from
the "softening" of the shape of $f_0$ as the temperature rises.
The widening of the Fermi distribution then implies significant
shift of the chemical potential as well as the progressive
activation of different types of electronic states in the
transport.

Our parametrization of $\rho(T)$ and $S(T)$ gives the following
values for the ratios: $\sigma_{\rm w}/\sigma\rm_n = 2.8$, and
$g_{\rm n}/g\rm_w=20$. The half-width of the narrow portion of the
DOS is $W\rm_n=60\,K$. The band filling at zero temperature is
slightly below the center of the narrow portion of DOS,
$\mu_0=-6.6\rm\,K$. The off-center shift is needed to explain the
finite TEP. At a finite temperature, the chemical potential is
calculated from the requirement that the number of particles stays
thermally independent. This set of parameters, as shown in Figure
\ref{model}, reproduces the temperature dependence of both the
resistivity and TEP over a wide temperature range, in good
accordance with experimental data. The value of $g_{\rm n}/g\rm_w$
is of the order of what has been calculated from the specific heat
measurements.\cite{JapSpecHeat,Markus} The temperature of the
pronounced maximum in $S(T)$  corresponds to the value of $W_n$.

The considerations of the microscopic sources of the values of
$g_n/g_w$, $\sigma_w$, and $\sigma_n$ are beyond the scope of this
paper. However, even at the present level we learn much about the
nature of the electronic states. Firstly, it is somewhat
surprising that the model does not require any separate
temperature dependencies for the parameters $\sigma\rm_n$ and
$\sigma\rm_w$. In fact, for the same model to reproduce both
$S(T)$ and $\rho(T)$ one condition is that this temperature
dependence be negligible. A sizeable temperature dependence of
$\sigma\rm_n$ and $\sigma\rm_w$ parameters would affect directly
the $\rho(T)$, whereas the additional temperature dependent
factors would cancel in the expression for $S(T)$. Thus most of
the temperature dependence comes exclusively from the existence of
two distinct parts in the electronic spectrum. Second observation
is linked to the strong increase of the resistivity in the low
temperature range, which implies that the rise in temperature
renders the charge carriers propagation more difficult. This is
contrary to what would happen if the low energy electronic states
were localized in space and it was the delocalized states that
became more populated as the temperature increased. The model
suggests that the mean free path $l_n$, for the states in the
enhanced part of DOS, is greater than the one related to the wide
part of the DOS, $l_w$. This is the consequence of the parameters
$\sigma_n$ and $\sigma_w$ being of similar order of magnitude,
\emph{i.e.} $v_n\,l_n\sim v_w\,l_w$, and a rather natural
assumption that the velocity in the wide portion of the electronic
spectrum is significantly larger than in the narrow part, $v_w \gg
v_n$. Good spatial coherence of the low lying states rules out a
bad metal or a localized transport limit. As the effective single
particle states close to $E\rm_F$ are spatially coherent, one may
speak of an effective, renormalized electronic dispersion at low
temperature, which is not destroyed by the weak residual
interaction. This situation, where the renormalization is strong
while the effective electronic band picture is preserved, is often
encountered in in heavy fermion systems.\cite{HottREVIEW}

The pressure dependence observed in the experiments may be
transferred into the pressure dependence of the model parameters.
Experiments unambiguously suggest that the pressure mostly affects
the low energy electronic spectrum. The model parameters related
to that part of the spectrum are the low energy DOS, $g\rm_n$, and
the scattering parameter $\sigma\rm_n$. The rattling of potassium
atom is supposed to weaken under pressure. This should mostly be
reflected in the gradual decrease of the DOS enhancement around
the $E\rm_F$. Indeed, reducing $g\rm_n$ by $20\%$ qualitatively
reproduces the observed shifts, both in the resistivity and TEP,
under the maximum pressure of $2.3\rm\,GPa$.

One final comment is due on the relation between the model and the
reported, almost temperature-independent magnetic susceptibility
$\chi$.\cite{Markus} It should be noted that a dynamically formed
enhancement in the DOS originating from the electron-phonon
interaction should generally not show up in $\chi$. The DOS
enhancement near the $E\rm_F$ forms separately for spin-up and
spin-down electrons. Hence, it cannot be regarded as a construct
that would stiffly move in energy in opposite directions for
spin-down and spin-up electrons when the magnetic field is
applied. The $E\rm_F$ is the same for spin-up and spin-down
electrons in a spin polarized system, therefore the DOS
enhancement should not move at all. As expected, no enhancement
was observed in the electronic energy as a result of the applied
magnetic field.\cite{Markus}

To conclude, we have studied the transport properties of $\rm
KOs_2O_6$ under pressures up to $2.3 \rm\,GPa$. A strong evidence
for a narrow enhancement in the DOS around the $E\rm_F$ comes from
the TEP data, which shows a non-metallic behavior persisting to
very high temperatures. A simple model is able to account for the
unusual features of both the thermopower and the resistivity over
a wide temperature range. We infer that the pressure behavior of
the transport coefficients is mainly influenced by the decrease in
rattling of the K ions.

We gratefully acknowledge discussions with I. Batisti\'c. This
work was supported by the Swiss National Science Foundation
through the NCCR pool MaNEP, and by the SCOPES Project No.
IB7320-111044. E.T. also acknowledges the support by the Ministry
of Science of Croatia (MZOS, Grant No. 035-0352826-2847).


\begin{thebibliography}{1}


\bibitem{KOSOdiscSC} S. Yonezawa, Y. Muraoka, Y. Matsushita, and Z. Hiroi, J. Phys.: Condens. Matter \textbf{16}, L9 (2004).

\bibitem{KOSOunprecedentedSC} Z. Hiroi, S. Yonezawa, and Y. Muraoka, J. Phys. Soc. Jpn. \textbf{73}, 1651 (2004).

\bibitem{AokiREV} H. Aoki, J. Phys.: Condens. Matter \textbf{16}, V1 (2004).

\bibitem{PickettRattling} J. Kune\v{s}, T. Jeong, and W. E. Pickett, Phys. Rev. B \textbf{70}, 174510 (2004).

\bibitem{RbSC} S. Yonezawa, Y. Muraoka, Y. Matsushita, and Z. Hiroi, J. Phys. Soc. Jpn. \textbf{73}, 819 (2004).

\bibitem{CsSC} S. Yonezawa, Y. Muraoka, and Z. Hiroi, J. Phys. Soc. Jpn. \textbf{73}, 1655 (2004).

\bibitem{JapSpecHeat} Z. Hiroi, S. Yonezawa, J.-I. Yamaura, T. Muramatsu, and Y. Muraoka,
J. Phys. Soc. Jpn. \textbf{74}, 1682 (2005).

\bibitem{Markus} M. Br\"uhwiler, S. M. Kazakov, J. Karpinski, and B. Batlogg, Phys. Rev. B \textbf{73}, 094518 (2006).

\bibitem{SanizBAND}  R. Saniz, J. E. Medvedeva, L.-H. Ye, T. Shishidou, and A. J. Freeman, Phys. Rev. B \textbf{70}, 100505(R) (2004).

\bibitem{egApexOxigen} A. R. Bishop, D. Mihailovic, and J. Mustre de León, J. Phys.: Condens. Matter \textbf{15},
L169 (2003).

\bibitem{vanHove} W. E. Pickett, Rev. Mod. Phys. \textbf{61}, 433 (1989);
K. Gofron, J. C. Campuzano, A. A. Abrikosov, M. Lindroos, A.
Bansil, H. Ding, D. Koelling, and B. Dabrowski, Phys. Rev. Lett.
\textbf{73}, 3302 (1994);
 A. Ino, C. Kim, M. Nakamura, T. Yoshida, T. Mizokawa, A. Fujimori, Z.-X. Shen, T. Kakeshita, H. Eisaki, and S. Uchida, Phys. Rev. B
\textbf{65}, 94504 (2002).

\bibitem{hpPRL}  T. Muramatsu, N. Takeshita, C. Terakura, H. Takagi,
Y. Tokura, S. Yonezawa, Y. Muraoka, and Z. Hiroi, Phys. Rev. Lett. \textbf{95}, 167004 (2005).

\bibitem{crystalgrowth}  G. Schuck, S. M. Kazakov, K. Rogacki, N. D. Zhigadlo, and J. Karpinski, Phys. Rev. B \textbf{73}, 144506 (2006).

\bibitem{NevenCdReO}N. Bari\v{s}i\'c, L. Forr\'o, D. Mandrus, R. Jin, J. He, and P. Fazekas, Phys. Rev. B \textbf{67}, 245112 (2003).

\bibitem{Ziman} J.M. Ziman, \emph{Electrons and phonons} (Oxford University Press,
1972), $\S$ 9.11.

\bibitem{Goodenough}J.-S. Zhou and J.B. Goodenough, Phys. Rev. B \textbf{51}, 3104 (1995).

\bibitem{CsRbSpecHeat} Z. Hiroi, S. Yonezawa, T. Muramatsu, J.-I. Yamaura, and Y. Muraoka,
J. Phys. Soc. Jpn. \textbf{74}, 1255 (2005).

\bibitem{KondoRattler} T. Hotta, Phys. Rev. Lett. \textbf{96}, 197201 (2006);
                  K. Hattori, Y. Hirayama, and K. Miyake,  J. Phys. Soc. Jpn.
                  \textbf{14}, 3306 (2005).

\bibitem{HottREVIEW} R. Hott, R. Kleiner, T. Wolf, and G. Zwicknagl, in A. Narlikar "Frontiers in
Superconducting Materials", Springer (2005), cond-mat/0408212


\end{thebibliography}
\end{document}